\documentstyle[aps,epsf]{revtex}
\begin{document}
\title{Reheating and turbulence}
\author{Mariana Gra\~{n}a$^{\dagger }$ and Esteban Calzetta$\ddagger $}
\address{$^{\dagger }$Department of Physics, University of \\
California at Santa Barbara, CA 93106, USA\\
$^{\ddagger }$Physics Department, University of Buenos Aires, Argentina}
\maketitle

\begin{abstract}
We show that the ''turbulent'' particle spectra found in numerical
simulations of the behavior of matter fields during reheating admit a simple
interpretation in terms of hydrodynamic models of the reheating period. We
predict a particle number spectrum $n_{k}\propto k^{-\alpha }$ with $\alpha
\sim 2$ for $k\rightarrow 0.$ Principal PACS No 98.80.Cq; additional PACS
nos: 04.62.+v, 11.10.Wx, 47.27
\end{abstract}

\section{Introduction}

The reheating period in the Early Universe stands out as a challenge to
theorists due to the close interrelationship of nonlinear and gravitational
phenomena in its unfolding (see \cite{paper0,paper1,paper2,paper2b,paper3},
henceforth ''papers 1-5''). The observation that, due to the high occupation
numbers produced during preheating, most of the physics of reheating may be
understood in terms of the behavior of nonlinear classical waves\cite
{paper0,Son} has been the key to sustantial progress. The authors of papers
1-5 have undertaken systematic numerical simulations of the behavior of
matter fields during reheating, finding that the reheating period is
actually composed of three consecutive phases: an early one or preheating,
where the dominant effect is the parametric amplification of matter fields
out of the dynamical inflaton and gravitational backgrounds \cite{preheating}%
, an intermediate stage where the dominant phenomenon is the redistribution
of energy among matter field modes through rescattering (in the terminology
of papers 1-5), and a final stage where thermal equilibrium sets in. During
the intermediate stage, spectra of occupation numbers for the matter fields
reduce to simple power laws both in the infrared and ultraviolet limits. As
noted in papers 1-5, this behavior suggests a connection between the physics
of reheating and the phenomena of weak turbulence \cite{Zakharov}, but to
the best of our knowledge no theoretical prediction for the exponents
involved is available. Our goal is to provide these theoretical estimates.

In this paper we shall follow this same trend of ideas, by observing that,
from the macroscopic point of view, a stochastic ensemble of classical waves
may be described by a conserved energy momentum tensor subject to the Second
Law of thermodynamics. There is therefore an equivalent fluid description,
consisting of a fluid whose energy momentum tensor and equation of state
reproduce the observed ones for the microscopic fluctuations. Solving the
dynamics of this equivalent fluid yields answers to all relevant questions
concerning the behavior of the actual fluctuations.

An immediate consequence of energy momentum conservation and the Second Law
is that when velocities are low, the phenomenological fluid may be described
within the Eckart theory of dissipative fluids \cite{Weinberg} (for an
analysis of the limitations of Eckart's theory see \cite{Geroch}). It
follows that it obeys a continuity equation and a curved space time Navier -
Stokes one. The ''turbulent ''spectra found in numerical simulations
correspond to the self-similar solutions discussed long ago by Chandrasekhar 
\cite{Chandra}. They appear in the discussion of decaying turbulence (which
is the case relevant to cosmology), as opposed to turbulence driven by some
external means. The Chandrasekhar solutions are built on the Heisenberg
closure hypothesis \cite{Hei48} (see \cite{Batchelor,UFrisch,McComb} for a
general discussion of turbulence). They were generalized to Friedmann -
Robertson - Walker (FRW) backgrounds by Tomita {\it et al }\cite{Tomita}.
These solutions agree with the Kolmogorov 1941 theory in the inertial range 
\cite{McComb}, failing to reproduce observations for very small eddies.
Fortunately, we are most interested in the opposite limit of very large
eddies, where it is trustworthy (we wish to point out that the applicability
of Kolmogorov's spectrum to large scale turbulence should not be taken for
granted \cite{GCPRL}). With minor adjustments, Tomita's analysis of
turbulence decay in FRW space times also provides a solution to the
evolution of our equivalent fluid.

The rest of the paper is organized as follows. In next section we provide a
brief summary of hydrodynamics in flat and expanding universes, in order to
set up the language for the rest of the paper, and introduce the self
similar solutions. In section III we proceed to discuss the equivalent fluid
description of field fluctuations, and how to extract the particle spectrum
therefrom. In section IV we place the self-similar solutions in the context
of reheating. We state our main conclusions in the final section. We provide
a rough estimate of the shear viscosity during reheating in the Appendix.

\section{Hydrodynamic flows}

\subsection{Flows in flat space time}

The equations governing the dynamics of a fluid in local thermodynamic
equilibrium are the continuity and Navier-Stokes ones, which, in the case of
flat space time, read:

\begin{equation}
\frac{\partial \rho }{\partial t}+\left( {\bf U\cdot }\nabla \right) \rho =0
\label{contpl}
\end{equation}

\begin{equation}
\frac{\partial {\bf U}}{\partial t}+\left( {\bf U\cdot }\nabla \right) {\bf %
U=-}\frac 1\rho \nabla p+\nu \nabla ^2{\bf U}  \label{NSpl}
\end{equation}
where we have assumed incompressibility, valid when typical velocities are
much smaller than the sound velocity; $\nu =\eta /\rho $ is the kinematic
shear viscosity. The transition from laminar to turbulent motion can be
universally described by the dimensionless ''Reynolds'' number:

\begin{equation}
R=\frac{UL}\nu
\end{equation}

where $U$ is a typical velocity and $L$ a typical length scale. This number
represents the order of magnitude of the ratio of the inertial to the
viscous term. Low Reynolds numbers correspond to laminar motion, while high
ones suggest turbulent behavior.

In general, the velocity profile displays variations in space and time. This
implies that the flow must be described probabilistically. Thus, each
quantity involved in (\ref{contpl}-\ref{NSpl}) is divided in its mean value
and a fluctuation from it; for example, we write ${\bf U}=\bar U+u$, where $%
u $ stands for the fluctuating part of the velocity. In the case where
motion is isotropic, the mean value $\bar U$ for the velocity must be zero,
since otherwise there would be a preferred direction.

To analyze the system's behavior, we define the two-point one-time
correlation function for the velocity:

\begin{equation}
R_{ij}({\bf x,x}^{\prime },t)=\left\langle u_i({\bf x,}t)u_j({\bf x}^{\prime
}{\bf ,}t)\right\rangle  \label{Rij}
\end{equation}
In the case of homogeneous and isotropic motion, this correlation must be
only a function of the time $t$ and the distance between ${\bf x}$ and ${\bf %
x}^{\prime }$, i.e. $R_{ij}({\bf x,x}^{\prime },t)=R_{ij}(r,t)$, where $%
r=\left| {\bf x-x}^{\prime }\right| .$ Observe that $R_{ii}(0,t)$ (summation
over repeated indices must be understood) is twice the average energy
density of the flow at time $t$. From (\ref{NSpl}) we obtain the equation
that this correlation must obey, namely:

\begin{equation}
\frac \partial {\partial t}R_{ij}(r,t)=T_{ij}(r,t)+P_{ij}(r,t)+2\nu \nabla
^2R_{ij}(r,t)  \label{ecRij}
\end{equation}
where 
\begin{equation}
P_{ij}(r,t)=\frac 1\rho \left( \frac \partial {\partial r_i}\left\langle p( 
{\bf x,}t{\bf )}u_j({\bf x}^{\prime },t)\right\rangle -\frac \partial {
\partial r_j}\left\langle p({\bf x}^{\prime },t)u_i({\bf x},t)\right\rangle
\right)  \label{Pij}
\end{equation}
and

\begin{equation}
T_{ij}(r,t)=\frac \partial {\partial r_k}\left\langle u_i({\bf x},t)u_k({\bf %
x},t)u_j({\bf x}^{\prime },t)-u_i({\bf x},t)u_k({\bf x}^{\prime },t)u_j({\bf %
x}^{\prime },t)\right\rangle  \label{Tij}
\end{equation}
The tensor $T_{ij}$ comes form the inertia term in Navier-Stokes equation
and, as it involves a product of third order in the velocity, reflects the
fact that there is not a close set of equation for the correlations of
successive orders but there is a hierarchy of equations instead. The problem
of closing that hierarchy is known as the ''moment closure problem'' \cite
{Schilling}. Let us call $\Phi _{ij}(k,t)$ the Fourier transform of $%
R_{ij}(r,t)$. Then the energy density becomes

\[
\frac 12R_{ii}(0,t)=\int E(k,t)\;dk, 
\]
where

\begin{equation}
E(k,t)=\frac 12\int \Phi _{ii}({\bf k},t)\ k^2\ d\Omega ({\bf k)}  \label{ET}
\end{equation}

is the energy density stored in eddies of size $k^{-1}.$ Defining $\Gamma
_{ij}$ as the Fourier transform of $T_{ij}$, we obtain from (\ref{ecRij})
the equation of balance of the energy spectrum: 
\begin{equation}
-\frac \partial {\partial t}E(k,t)=T(k,t)+2\nu k^2E(k,t)  \label{ecesp}
\end{equation}
where 
\begin{equation}
\text{\qquad }T(k,t)=-\frac 12\int \Gamma _{ii}({\bf k},t)\ k^2\ d\Omega ( 
{\bf k)}  \label{ET1}
\end{equation}

The inertia term $T(k,t)$ is the one that contains the mode-mode
interaction, and its effect is to drain energy form the more energetic modes
-typically the bigger ones- to the ones where there is major viscous
dissipation -the smaller ones-.

\subsection{Flows in expanding universes}

For a curved space-time, in particular a Friedmann - Robertson - Walker
(FRW) Universe with zero spatial curvature ($ds^2=-dt^2+a^2(t)\left(
dx^2+dy^2+dz^2\right) $), the generalization of the above arguments has been
considered by many authors \cite
{vonWeizsacker,Nariai,SMT,OzernoiyChibisov,NariaiyTanabe}. We follow Tomita 
{\it et al.}'s analysis \cite{Tomita}, in which they obtain the solution for
the energy spectrum in the case of homogeneous, isotropic and incompressible
turbulence.

In a generic space time, we describe fluid flow from the energy density $%
\rho $, pressure $p$ and four velocity $U$. The symmetries of the FRW
solution suggest using instead the commoving three velocity $u^i=U^i/U^0;$
if $U^i\ll U^0$ the flow is non relativistic, and if $\nabla {\bf u=}0,$ it
is incompressible (${\bf u}=(u^1,u^2,u^3)$). Later on, we shall also use the
physical three velocity $v=a(t)u.$

The corresponding continuity and Navier-Stokes equations for a
Robertson-Walker background are obtained by the condition of conservation of
the energy-momentum tensor \cite{Weinberg}. For a non relativistic
incompressible fluid, with shear viscosity $\eta =\nu \left( p+\rho \right) $
(but no bulk viscosity), these reduce to:

\begin{equation}
\frac{\partial \rho }{\partial t}+3\frac{\dot a}a\left( p+\rho \right) =0
\label{contcur}
\end{equation}
\begin{equation}
\frac{\partial {\bf u}}{\partial t}+\left[ \left( {\bf u\cdot }\nabla
\right) +\frac{\partial \ln \left( (p+\rho )a^5\right) }{\partial t}\right] 
{\bf u=-}\frac{\nabla p}{a^2\left( p+\rho \right) }+\frac 1{a^2}\nu \nabla
^2 {\bf u}  \label{NScur}
\end{equation}
where we have assumed that $p+\rho $ depends only on time. For the physical
three velocity ${\bf v}$, the corresponding Navier-Stokes equation reads: 
\begin{equation}
\frac{\partial {\bf v}}{\partial t}+\left[ \frac 1a\left( {\bf v\cdot }
\nabla \right) +\frac{\partial \ln \left( (p+\rho )a^4\right) }{\partial t}
\right] {\bf v=-}\frac{\nabla p}{a\left( p+\rho \right) }+\frac 1{a^2}\nu
\nabla ^2{\bf v}  \label{NScurv}
\end{equation}
In obtaining (\ref{contcur})-(\ref{NScurv}) we have neglected possible
perturbations to the FRW metric. The corresponding equations considering
fluctuations in the metric ($g_{\mu \nu }=g_{\mu \nu }^0+h_{\mu \nu }$) have
been obtained by Weinberg \cite{Weinberg}. The continuity equation is not
corrected by gravitational perturbations, while in the Navier-Stokes
equation the metric fluctuations appear explicitly only within the shear
viscosity term. It can be demonstrated that these terms involving metric
fluctuations are negligible for scales that are inside the horizon \cite{GC}
. For scales bigger than the Hubble radius, since dissipation through
viscosity is not effective anyway, we may still use the unperturbed Navier -
Stokes equation.

The operation of Fourier transforming in the case of a Robertson-Walker
cosmology is done in terms of commoving wave-numbers. In doing so, the
following equation for the energy spectrum is obtained: 
\begin{equation}
-\frac \partial {\partial t}E(k,t)=T(k,t)+2\left\{ \frac{\nu k^2}{a^2}+\frac{
\partial \ln \left( (p+\rho )a^4\right) }{\partial t}\right\} E(k,t)
\label{ecespcur}
\end{equation}
where the relationship between $E(k,t)$ and $\Phi _{ij}(k,t)$ as well as
between $T(k,t)$ and $\Gamma _{ij}(k,t)$ is the same as that for a flat
space time, if we define $R_{ij}$ and $T_{ij}$ from correlations of physical
quantities, as follows: 
\begin{eqnarray}
R_{ij}(r,t) &=&a^2\left\langle u_i({\bf x,}t)u_j({\bf x+r},t)\right\rangle
,\   \label{ETcur} \\
\ T_{ij}(r,t) &=&a^2\frac \partial {\partial r_k}\left( \left\langle u_i( 
{\bf x},t)u_k({\bf x},t)u_j({\bf x+r},t)\right\rangle -\left\langle u_i({\bf %
x},t)u_k({\bf x+r},t)u_j({\bf x+r},t)\right\rangle \right)
\end{eqnarray}

\subsection{Self similar flows in flat and expanding universes}

As we have seen in the previous section, the key element in the description
of the flow is the energy spectrum $E(k)$ (Eq.(\ref{ET})), which is the
solution of the balance equation (Eq.(\ref{ecesp})). In it, the right hand
side contains the viscous dissipation as well as the inertial force $T(k,t$.
The overall effect of this term is to transfer energy from a given scale to
smaller ones through mode - mode coupling; thus it is natural to model the
action of the inertia term as a source of viscous dissipation, where the
effective turbulent viscosity for a given mode depends on the motion of all
smaller eddies \cite{UFrisch}. By providing closure, that is, writing this
effective viscosity in terms of the spectrum itself, a closed evolution
equation for $E(k)$ is obtained. Concretely, Heisenberg \cite{Hei48}
proposed the ansatz

\begin{equation}
\int_0^kT(k^{\prime },t)\ dk^{\prime }=2\nu (k,t)\int_0^kE(k^{\prime },t)\
k^{\prime \ 2}\ dk^{\prime }  \label{Heis1}
\end{equation}
where

\begin{equation}
\nu (k,t)=A_{flat}\int_k^\infty \sqrt{\frac{E(k^{\prime },t)}{k^{\prime 3}}}
\ dk^{\prime }  \label{Heis2}
\end{equation}
and $A_{flat}$ is a dimensionless constant. With this hypothesis (known as
the Heisenberg hypothesis) as the solution to the closure problem,
Chandrasekhar \cite{Chandra} has obtained the energy spectrum for decaying
turbulence, assuming that there is a stage in the decay where the bigger
eddies have sufficient amount of energy to maintain an equilibrium
distribution, thus requiring that the solution for the spectrum should be
self similar. With this consideration into account he obtained an energy
spectrum:

\begin{equation}
E(k,t)=\frac 1{A_{flat}^2k_0^3t_0^2}\sqrt{\frac{t_0}t}F\left( \frac{k\sqrt{t}
}{k_0\sqrt{t_0}}\right) \qquad  \label{flatspacelim}
\end{equation}

where $k_0$ and $t_0$ are initial conditions (namely, the wave number
corresponding to the bigger eddy and its typical time of evolution). The
function $F$ obeys the equation

\begin{equation}
\frac 14\int_0^x\left[ F(x^{\prime })-x^{\prime }\frac{dF(x^{\prime })}{
dx^{\prime }}\right] dx^{\prime }=\left\{ \nu k_0^2t_0+\int_x^\infty \frac{%
\sqrt{F(x^{\prime })}}{x^{\prime \ 3/2}}dx^{\prime }\right\}
\int_0^xF(x^{\prime })x^{\prime \ 2}dx^{\prime }  \label{ecF}
\end{equation}
which predicts a Kolmogorov type behavior for an inviscid fluid ($%
R\rightarrow \infty $, $R=\left( \nu k_0^2t_0\right) ^{-1}$) in the
ultraviolet limit:

\begin{equation}
F(x)\rightarrow const\ x^{-5/3}\ (\nu =0\ ,\ x\rightarrow \infty )
\end{equation}

While for nonzero viscosity:

\begin{equation}
F(x)\rightarrow const\ x^{-7}\ (\nu \neq 0\ ,\ x\rightarrow \infty )
\end{equation}

In the infrared limit, $F$ has the universal behavior $F(x)=4x$ $(x\ll 1)$,
and thus we find a linear energy spectrum for $k\sqrt{t}\ll k_0\sqrt{t_0}.$

Chandrasekhar's self similar solutions are easily generalized to flows in
expanding Universes. The dependence on time and wave-number for the self
similar energy spectrum is \cite{Tomita}

\begin{equation}
E(k,t)=\frac{1}{2}v_{t}^{2}\left( t\right) \lambda (t)F(\lambda k)
\label{speccur1}
\end{equation}
where $\lambda $ and $v_{t}$ are respectively the Taylor's microscale and an
average turbulent velocity, defined as:

\begin{equation}
\lambda ^2(t)\equiv 5\frac{\int E(k,t)\ dk}{\int E(k,t)\ k^2\ dk}\ \qquad 
\frac 12v_t^2(t)\equiv \int E(k,t)\ dk  \label{lambdayvt}
\end{equation}

The second equation implies the normalization condition

\begin{equation}
\int_0^\infty F(x^{\prime })dx^{\prime }=1  \label{normalization}
\end{equation}

To obtain a self similar flow reducing to Eq. (\ref{flatspacelim}) in the
flat space limit, we must require the time evolution laws: 
\begin{equation}
\lambda ^2(t)=\lambda _i^2+10\int_{t_i}^t\frac \eta {(p+\rho )a^2}\ dt\qquad
v_t=v_{ti}\left( \frac{(p+\rho )_ia_i^4}{(p+\rho )a^4}\right) \frac{\lambda
_i}{\lambda \left( t\right) }  \label{l(t)yv(t)}
\end{equation}

The equation which determines de function $F(\lambda k)$ in (\ref{speccur1})
turns out to be

\begin{equation}
\int_0^x\left[ F(x^{\prime })-x^{\prime }\frac{dF(x^{\prime })}{dx^{\prime }}
\right] dx^{\prime }=\left\{ \frac 25+A\int_x^\infty \frac{\sqrt{F(x^{\prime
})}}{x^{\prime \ 3/2}}dx^{\prime }\right\} \int_0^xF(x^{\prime })x^{\prime \
2}dx^{\prime }  \label{curvedchandra}
\end{equation}

where $A$ is a constant. This equation has the same structure as in flat
space time, Eq. (\ref{ecF}), which means that assuming Heisenberg's
hypothesis the spectrum is linear in $k$ for length scales much bigger than
the Taylor's microscale.

Observe that for flat space time, the proportionality between the integral
up to a certain wave number $k$ of the inertia and the viscous forces is
given by (\ref{Heis1}) and (\ref{Heis2}). In the case of a FRW space time,
the autosimilar solution required by Tomita {\it et al.} (\ref{speccur1})
needs a time dependent $\nu _{curv}$ in eq. (\ref{Heis1}), defined by the
analog of (\ref{Heis2}) with $A_{flat}$ replaced by $A_{curv}=5A\eta a^{2}$.

\subsection{Solving for the spectrum}

Let us analyze in more detail the solutions of eq. (\ref{curvedchandra}). We
assume the normalization eq. (\ref{normalization}). By taking the $%
x\rightarrow \infty $ limit in eq. (\ref{curvedchandra}) we find

\begin{equation}
\int_0^\infty F(x^{\prime })x^{\prime 2}dx^{\prime }=5
\label{normalization2}
\end{equation}

Taking a derivative of eq. (\ref{curvedchandra}) we get

\begin{equation}
1-x\frac{F^{\prime }}F=x^2\left\{ \frac 25+A\left( G-H\right) \right\}
\label{differential}
\end{equation}

where

\begin{equation}
G=\int_x^\infty \frac{\sqrt{F(x^{\prime })}}{x^{\prime \ 3/2}}dx^{\prime }
\label{gfunction}
\end{equation}

\begin{equation}
H=\frac 1{\sqrt{Fx^7}}\int_0^xF(x^{\prime })x^{\prime \ 2}dx^{\prime }
\label{hfunction}
\end{equation}

Let us consider first the $x\rightarrow 0$ limit. Assume $F\propto x^{\alpha
}$. The left hand side of eq. (\ref{differential}) tends to a finite limit $%
1-\alpha $. On the right hand side, $G$ and $H$ behave as $x^{\left( \alpha
-1\right) /2}$, so if $\alpha >0,$ this side goes to zero. We must therefore
have $\alpha =1$, and

\begin{equation}
F\sim Cx,\qquad x\rightarrow 0  \label{fsmall}
\end{equation}

were $C$ is some constant.

In the $x\rightarrow \infty $ limit, assume again a power law behavior $%
F\propto x^{-\beta }$. Now $G\rightarrow 0$, so we must have $H\rightarrow
2/5A.$ From eq. (\ref{normalization2}) we know that in this limit $H\sim 5/ 
\sqrt{Fx^7}$, so we must have $\beta =7$ and

\begin{equation}
F\sim \left( \frac{25A}2\right) ^2x^{-7},\qquad x\rightarrow \infty
\label{flarge}
\end{equation}

Taking into account both limiting behaviors and eq. (\ref{normalization}),
the function $F$ may be approximated as

\begin{equation}
F\left[ x\right] =\frac{x}{\left[ \alpha +\beta x^{4}\right] ^{2}},\qquad
\alpha =\left[ \frac{25A\pi ^{2}}{128}\right] ^{1/3}\qquad \beta =\frac{2}{
25A}  \label{fapprox}
\end{equation}

\section{Equivalent fluid for field fluctuations}

After establishing the basic necessary notions for the description of
hydrodynamic flows, our goal is to associate an equivalent fluid description
to field fluctuations, and to derive the particle spectrum therefrom. Our
first step is to obtain the energy density, pressure and velocity of this
fluid as functionals of the quantum state of the field.

For simplicity, we shall consider the theory of a single, self-interacting
scalar field $\phi $, minimally coupled to gravity. The action is

\begin{equation}
S=\int d^{4}x\,\sqrt{-g}\left\{ \left( \frac{-1}{2}\right) \partial _{\mu
}\phi \partial ^{\mu }\phi -V\left[ \phi \right] \right\}
\end{equation}

where $V\left[ \phi \right] $ ia renormalized effective potential. The
energy-momentum tensor is associated to the Heisenberg operator

\begin{equation}
T_{Q}^{\mu \nu }=\frac{\left( -2\right) }{\sqrt{-g}}\frac{\delta S}{\delta
g_{\mu \nu }}=\partial ^{\mu }\phi \partial ^{\nu }\phi -g^{\mu \nu }\left\{
\left( \frac{1}{2}\right) \partial _{\rho }\phi \partial ^{\rho }\phi
+V\left[ \phi \right] \right\}
\end{equation}

The macroscopic behavior of the field, however, may be described in terms of
a c-number energy-momentum tensor

\begin{equation}
T^{\mu \nu }=T_{C}^{\mu \nu }+T_{S}^{\mu \nu }
\end{equation}

where

\begin{equation}
T_{C}^{\mu \nu }=\left\langle T_{Q}^{\mu \nu }\right\rangle _{Q}
\end{equation}

and $T_S^{\mu \nu }$ is a stochastic component with zero mean and
self-correlation

\begin{equation}
\left\langle T_{S}^{\mu \nu }T_{S}^{\rho \sigma }\right\rangle _{S}=\frac{1}{
2}\left\langle \left\{ T_{Q}^{\mu \nu },T_{Q}^{\rho \sigma }\right\}
\right\rangle _{Q}-T_{C}^{\mu \nu }T_{C}^{\rho \sigma }
\end{equation}

In these equations, $\left\langle {}\right\rangle _S$ denotes a stochastic
average, while $\left\langle {}\right\rangle _Q$ is the average with respect
to the quantum state of the field. Following Landau, we define the fluid
four-velocity $U^\mu $ and energy density $\rho $ as the (only) time-like
eigenvector of $T^{\mu \nu }$ and (minus) its corresponding eigenvalue

\begin{equation}
T^{\mu \nu }U_{\nu }=-\rho U^{\mu }
\end{equation}

Introducing the pressure $p=p\left( \rho \right) $ as given by the
equilibrium equation of state (our theory does not have a conserved particle
number current, and therefore the equation of state is barotropic), we may
decompose

\begin{equation}
T^{\mu \nu }=\rho U^{\mu }U^{\nu }+p\Delta ^{\mu \nu }+\tau ^{\mu \nu }
\end{equation}

where $\Delta ^{\mu \nu }=g^{\mu \nu }+U^\mu U^\nu $ and by construction $%
\tau ^{\mu \nu }U_\nu =0$. Since $\tau ^{\mu \nu }$ vanishes by definition
in the equilibrium state, it may be parametrized in terms of deviations from
equilibrium. Remaining within the so-called {\it first order formalism }\cite
{israel88,fof}, we may write

\begin{equation}
\tau ^{\mu \nu }=-\eta H^{\mu \nu }-\zeta U_{,\rho }^\rho \Delta ^{\mu \nu
};\qquad \eta ,\zeta \geq 0  \label{noneq}
\end{equation}

where

\begin{equation}
H^{\mu \nu }=\frac 12\Delta ^{\mu \lambda }\Delta ^{\nu \sigma }\left[
U_{\lambda ,\sigma }+U_{\sigma ,\lambda }-\frac 23\Delta _{\lambda \sigma
}U_{,\rho }^\rho \right]  \label{hmunu}
\end{equation}

and $\eta $ and $\zeta $ are the {\it shear }and {\it bulk} viscosity
coefficients, respectively.

Let us decompose each quantity in a mean component (denoted by a $C$
subscript) and a fluctuation (denoted by a $S$). If the quantum state shares
the symmetries of the FRW background, then $U_C^i=0$. Since $U^2=-1$ holds
identically (as oppossed to ''in the mean'') we must have

\begin{equation}
\left( U_{C}^{0}\right) ^{2}+\left\langle \left( U_{S}^{0}\right)
^{2}\right\rangle -a^{2}\left\langle U_{S}^{i}U_{S}^{i}\right\rangle _{S}=1
\end{equation}

\begin{equation}
2U_{C}^{0}U_{S}^{0}-a^{2}\left[ U_{S}^{i}U_{S}^{i}-\left\langle
U_{S}^{i}U_{S}^{i}\right\rangle _{S}\right] =0
\end{equation}

The second equation shows that $U_S^0$ is a higher order fluctuation with
respect to $U_S^i$. If we remain at linear order, then, we may approximate $%
U^0=U_C^0=1.$ Observing that all mean values are homogeneous and isotropic,
we see that $\tau ^{0i}$ is also a higher order fluctuation. We find

\begin{equation}
T^{0i}=T_{S}^{0i}=\left( \rho +p\right) _{C}U_{S}^{i}
\end{equation}

and therefore the velocity correlation

\begin{equation}
R^{ij}\left( \vec{r},t\right) =\frac{a^{2}\left( t\right) }{2\left( \rho
+p\right) _{C}^{2}\left( t\right) }\left\langle \left\{ T_{Q}^{0i}\left( 
\vec{r},t\right) ,T_{Q}^{0j}\left( 0,t\right) \right\} \right\rangle _{Q}
\end{equation}

This is the key equation linking the quantum and stochastic descriptions. To
estimate the velocity correlation, let us assume that, after integrating out
the hard modes, the soft modes of interest may be described in terms of
quasi free, long lived excitations with an effective mass $M^2\left(
t\right) .$ Then

\begin{equation}
T_{Q}^{0i}=\left( \frac{-1}{a^{2}\left( t\right) }\right) \partial _{t}\phi
\partial _{i}\phi
\end{equation}

The fluctuations are Gaussian to a very good approximation, and therefore

\begin{equation}
\left\langle \left\{ T_{Q}^{0i}\left( \vec{r},t\right) ,T_{Q}^{0j}\left(
0,t\right) \right\} \right\rangle _{Q}=\left( \frac{1}{a^{4}\left( t\right) }
\right) \left\{ \partial _{tj^{\prime }}^{2}G^{+}\partial _{it^{\prime
}}^{2}G^{+}+\partial _{ij^{\prime }}^{2}G^{+}\partial _{tt^{\prime
}}^{2}G^{+}+\left( \vec{r}\rightarrow -\vec{r}\right) \right\}
\end{equation}

where $G^{+}\left( x,x^{\prime }\right) $ is the positive frequency
propagator

\begin{equation}
G^{+}\left( \left( \vec{r},t\right) ,\left( 0,t\right) \right) =\left\langle
\phi \left( \vec{r},t\right) \phi \left( 0,t\right) \right\rangle _{Q},
\end{equation}

($\partial _{i,t}$ stand for derivatives with respect to the first argument,
while $\partial _{i^{\prime },t^{\prime }}$ stand for derivatives with
respect to the second argument of $G^{+}$). Let us decompose the soft field
into modes

\begin{equation}
\phi \left( \vec{r},t\right) =\int \frac{d^{3}k}{\left( 2\pi \right) ^{3}}
\,e^{i\vec{k}\vec{r}}\phi _{\vec{k}}\left( t\right)
\end{equation}

At any time $t_{0}$ we may introduce positive frequency adiabatic modes
defined by \cite{BirDv}

\begin{equation}
f_{k}\left( t\right) =\frac{1}{\sqrt{2\omega _{k}\left( t\right) }}{\rm exp}
\left\{ -i\int_{t_{0}}^{t}dt^{\prime }\,\omega _{k}\left( t^{\prime }\right)
\right\}
\end{equation}

where

\begin{equation}
\omega _{k}^{2}\left( t\right) =\frac{k^{2}}{a^{2}\left( t\right) }
+M^{2}\left( t\right)
\end{equation}

and decompose the mode amplitude $\phi _{\vec k}\left( t\right) $ into
positive and negative frequency components

\begin{equation}
\phi _{\vec{k}}\left( t\right) =f_{k}\left( t\right) A_{\vec{k}}\left(
t\right) +f_{k}^{*}\left( t\right) A_{-\vec{k}}^{\dagger }\left( t\right)
\end{equation}

\begin{equation}
\partial _{t}\phi _{\vec{k}}\left( t\right) =-i\omega _{k}\left( t\right)
\left\{ f_{k}\left( t\right) A_{\vec{k}}\left( t\right) -f_{k}^{*}\left(
t\right) A_{-\vec{k}}^{\dagger }\left( t\right) \right\}
\end{equation}

Let us define the spectrum

\begin{equation}
n_{k}\left( t\right) =\left\langle A_{\vec{k}}^{\dagger }\left( t\right) A_{%
\vec{k}}\left( t\right) \right\rangle _{Q}
\end{equation}

(because the quantum state is isotropic, the spectrum depends only on $k$)
and assume that

\begin{equation}
\left\langle A_{\vec{k}^{\prime }}\left( t\right) A_{\vec{k}}\left( t\right)
\right\rangle _{Q}=\left\langle A_{\vec{k}^{\prime }}^{\dagger }\left(
t\right) A_{\vec{k}}^{\dagger }\left( t\right) \right\rangle _{Q}=\left.
\left\langle A_{\vec{k}^{\prime }}^{\dagger }\left( t\right) A_{\vec{k}
}\left( t\right) \right\rangle _{Q}\right| _{\vec{k}^{\prime }\neq \vec{k}}=0
\end{equation}

(this happens, for example, if the different modes acquire random phases
through interaction with an environment). Then

\begin{equation}
G^{+}\left( \left( \vec{r},t\right) ,\left( 0,t^{\prime }\right) \right)
=\int \frac{d^{3}k}{\left( 2\pi \right) ^{3}}\,e^{i\vec{k}\vec{r}}\,\left\{
f_{k}\left( t\right) f_{k}^{*}\left( t^{\prime }\right) \left( 1+n_{k}\left(
t\right) \right) +f_{k}^{*}\left( t\right) f_{k}\left( t^{\prime }\right)
n_{k}\left( t\right) \right\}
\end{equation}

And

\begin{equation}
\partial _{t}G^{+}\left( \left( \vec{r},t\right) ,\left( 0,t^{\prime
}\right) \right) =\int \frac{d^{3}k}{\left( 2\pi \right) ^{3}}\,e^{i\vec{k} 
\vec{r}}\,\left( -i\omega _{k}\left( t\right) \right) \left\{ f_{k}\left(
t\right) f_{k}^{*}\left( t^{\prime }\right) \left( 1+n_{k}\left( t\right)
\right) -f_{k}^{*}\left( t\right) f_{k}\left( t^{\prime }\right) n_{k}\left(
t\right) \right\}
\end{equation}

Observe that only the vacuum part contributes in the coincidence limit $%
t^{\prime }\rightarrow t$. In the large occupation numbers regime we are
interested in, this is negligible, and we get

\begin{equation}
\left\langle \left\{ T_{Q}^{0i}\left( \vec{r},t\right) ,T_{Q}^{0j}\left(
0,t\right) \right\} \right\rangle _{Q}\sim \left( \frac{1}{a^{4}\left(
t\right) }\right) \left\{ \partial _{ij^{\prime }}^{2}G^{+}\partial
_{tt^{\prime }}^{2}G^{+}+\left( \vec{r}\rightarrow -\vec{r}\right) \right\}
\end{equation}

where

\begin{equation}
\partial _{ij^{\prime }}^{2}G^{+}=\frac{1}{3}\delta _{ij}\int \frac{d^{3}k}{%
\left( 2\pi \right) ^{3}}\,e^{i\vec{k}\vec{r}}\,\left( \frac{k^{2}}{\omega
_{k}\left( t\right) }\right) n_{k}\left( t\right)
\end{equation}

\begin{equation}
\partial _{tt^{\prime }}^{2}G^{+}=\int \frac{d^{3}k}{\left( 2\pi \right)
^{3} }\,e^{i\vec{k}\vec{r}}\,\omega _{k}\left( t\right) n_{k}\left( t\right)
\end{equation}

We may now write down the Fourier transform of the velocity self correlation

\begin{equation}
\Phi ^{ij}\left( k,t\right) =\frac{\delta ^{ij}}{3a^{2}\left( t\right)
\left( \rho +p\right) _{C}^{2}\left( t\right) }\int \frac{d^{3}p}{\left(
2\pi \right) ^{3}}\,\frac{\omega _{p}\left( t\right) }{\omega _{\left| \vec{
k }-\vec{p}\right| }\left( t\right) }\left| \vec{k}-\vec{p}\right|
^{2}n_{p}\left( t\right) n_{\left| \vec{k}-\vec{p}\right| }\left( t\right)
\end{equation}

In principle, this is an integral equation relating the spectrum to the
energy self-correlation. It may be simplified as follows. For small $p\leq k$%
, the integral reads

\begin{equation}
\left( \frac{k^{2}}{\omega _{k}\left( t\right) }\right) n_{k}\left( t\right)
\int \frac{d^{3}p}{\left( 2\pi \right) ^{3}}\,\omega _{p}\left( t\right)
n_{p}\left( t\right)  \label{smallp}
\end{equation}

while for large $p$ we get

\begin{equation}
\int \frac{d^{3}p}{\left( 2\pi \right) ^{3}}\,p^{2}n_{p}^{2}\left( t\right)
\label{largep}
\end{equation}

Since the spectra we are considering fall off much faster than $p^{-2}$ (or
even $p^{-3}$) at large $p,$ we may estimate that the contribution from eq. (%
\ref{smallp}) is larger than eq. (\ref{largep}); moreover, extending the
integral to the whole momentum space,

\begin{equation}
\int \frac{d^{3}p}{\left( 2\pi \right) ^{3}}\,\omega _{p}\left( t\right)
n_{p}\left( t\right) \sim \rho \left( t\right)
\end{equation}

and we get

\begin{equation}
\Phi ^{ij}\left( k,t\right) =\frac{\rho \left( t\right) \delta ^{ij}}{
3a^{2}\left( t\right) \left( \rho +p\right) _{C}^{2}\left( t\right) }\left( 
\frac{k^{2}}{\omega _{k}\left( t\right) }\right) n_{k}\left( t\right)
\end{equation}

Finally, we may relate the particle spectrum to the turbulent energy spectrum

\begin{equation}
E\left( k,t\right) =\frac{2\pi \rho \left( t\right) }{a^{2}\left( t\right)
\left( \rho +p\right) _{C}^{2}\left( t\right) }\left( \frac{k^{4}}{\omega
_{k}\left( t\right) }\right) n_{k}\left( t\right)  \label{fluid2quantum}
\end{equation}

\section{Reheating}

In the previous sections, we have analyzed on one hand self similar
turbulent flows in expanding universes, and on the other hand have given the
rule to translate the turbulent energy spectrum into a particle number
distribution. We must now show that the foregoing analysis is relevant to
plausible models of the reheating period, and use it to predict the likely
shape of the final particle spectrum.

At this point, it is convenient to be more precise about the model of
reheating we have in mind (for further details, see papers 1-5). We assume
that inflation is driven by an inflaton field $\Phi $ with an effective
potential which may be parameterized as $V\left( \Phi \right) \sim \lambda
_{inf}\Phi ^{4}.$ The self coupling $\lambda _{inf}\sim 10^{-14},$and at the
end of inflation $\Phi \sim m_{p}\sim 10^{19}GeV$ (in natural units). This
is the dominant contribution to the energy density, so it fixes the scale
for the Hubble constant $H$: $m_{p}^{2}H^{2}\sim V\left( \Phi \right) $.

During preheating, a large fraction (nearly all) of this energy is
transferred to the matter fields. These are described by an effective degree
of freedom $\phi $ which self interacts with an effective dimensionless
coupling constant $g.$ For simplicity, we shall model this self interaction
as a $g\phi ^{4}$ theory. At times $t_{0}$ at the end of preheating, matter
excitations are distributed with a smooth spectrum $n_{k}\sim Nf\left(
k/k_{0}\right) $, where $k$ is a comoving wave number and the characteristic
momentum $k_{0}\succeq H$ (meaning that the relevant modes are inside the
horizon). We wish to find the form of the spectrum as follows from the
assumption that its further evolution will be self-similar.

Let us estimate the energy density in the matter fields at $t_{0}$ as $\rho
_{0}\sim Nk_{0}^{4}\sim \lambda _{inf}\Phi ^{4}$. The self interaction of
the matter fields induces a mass $M^{2}$. We shall assume $k_{0}^{2}\geq
M^{2},$ in which case

\begin{equation}
M^{2}\sim g\int \frac{d^{3}k}{k}\;n_{k}\sim gNk_{0}^{2}\sim g\left( \rho
_{0}/k_{0}^{2}\right) .
\end{equation}

We shall assume that this is the dominant contribution to the matter
self-energy. The condition $k_{0}^{2}\geq M^{2}$ is equivalent to

\begin{equation}
gN\leq 1  \label{lambdan}
\end{equation}

(this equation will be important below). Observe that

\begin{equation}
\frac{M^{2}}{k_{0}^{2}}\sim \frac{g\rho _{0}}{k_{0}^{4}}\sim \left( \frac{g}{%
\lambda _{inf}}\right) \left( \frac{\rho _{0}}{k_{0}^{4}}\right) \left( 
\frac{V\left( \Phi \right) }{\Phi ^{4}}\right) \sim \left( \frac{g}{\lambda
_{inf}}\right) \left( \frac{H}{k_{0}}\right) ^{4}\left( \frac{m_{p}}{\Phi }%
\right) ^{4}
\end{equation}

Since $H/k_0$ is already less than one, this is not unduly restrictive on $g$
(see papers 1-5)$.$ Also observe that $M^2$ redshifts with the cosmological
evolution, so these estimates are not affected by Hubble flow.

Integrating equation (\ref{fluid2quantum}) over $k$, we find the mean
velocity in the equivalent fluid flow as

\begin{equation}
v_{ti}^{2}\sim \frac{\rho _{0}p_{0}}{\left( \rho _{0}+p_{0}\right) ^{2}}\leq
c_{s}^{2}
\end{equation}

where

\begin{equation}
c_{s}^{2}\sim \frac{k_{0phys}^{2}}{k_{0phys}^{2}+M^{2}}
\end{equation}

($k_{0phys}=k_{0}/a\left( t\right) $) is the speed of sound, so the flow may
be regarded as incompressible. The shear viscosity of the matter fields may
be estimated as (see appendix) \cite{Jeon}

\begin{equation}
\eta \sim \frac{k_{0}^{3}}{Ng^{2}}  \label{noneqvisc}
\end{equation}

leading to a kinematic viscosity

\begin{equation}
\nu \equiv \frac{\eta }{p+\rho }\sim \frac{1}{\left( gN\right) ^{2}k_{0}}
\end{equation}

With a typical velocity of order one, and a typical lenght of order $%
k_0^{-1},$ we get the Reynolds number $R\sim \left( gN\right) ^2\leq 1$
(cfr. eq. (\ref{lambdan})).

Of course, we do not have a microscopic justification for Heisenberg's
closure, so we shall take $A$ in eq. (\ref{curvedchandra}) as a free
parameter. As it turns out, agreement with the numerical results reported by
papers 1-5 is obtained for $A\sim 3,$ which, as expected, corresponds to
Reynolds numbers of order of one.

It's worth to point out that Chandrasekhar's solution with the function $F$
obeying eq. (\ref{curvedchandra}) is an exact solution that relies only on
Heisenberg's closure hypotesis eq. (\ref{Heis1}). So if we trust this
hypotesis, we can get the spectrum by solving eq. (\ref{curvedchandra}) for
any Reynolds number.

The distribution of occupation numbers is found from eq. (\ref{fluid2quantum}%
), where the energy spectrum is given by eq. (\ref{speccur1}). In the light
field limit $M^{2}\ll k_{0}^{2},$ we find $\omega _{k}\sim k$. Then eq. (\ref
{fsmall}) implies $n_{k}\sim k^{-2}$ for $k\rightarrow 0,$ and eq. (\ref
{flarge}) implies $k^{-10}$ for $k\rightarrow \infty $. This theoretical
prediction for the exponents involved is the main result of this paper.

In Fig. 1 we show the full particle spectrum based on the approximation eq. (%
\ref{fapprox}) for the function $F$. We have scaled the plot to make it
easiest to compare with papers 1-5. Momentum is measured in units of $%
a^{-1}10^{12}GeV$ (where $a$ is the scale factor) \cite{paper2b}, and we
have chosen $\lambda \sim a\,1.8\,\times 10^{-13}GeV^{-1}.$ Observing that
in the ultrarelativistic limit the speed of flow $v_t\sim 1$ and $p\sim \rho 
$, integrating the particle density times $k^3$ shows that the total energy
density in the flow is $\rho \sim 10^9\times \left( 10^{12}GeV\right) ^4$.
The equivalent black-body temperature is then somewhat less than $10^{15}GeV$%
, which is a reasonable value for the reheating era, and high enough to
justify the neglect of all masses.

The analysis leading to the Chandrasekhar solutions begins from the Navier -
Stokes equations for the fluid, which are equivalent in this context to
energy - momentum conservation. Therefore our model requires that it be
possible to assign to the fluid an independently conserved energy - momentum
tensor. This is not exactly the same as requiring that the homogeneous mode
has totally decayed, but it does mean that there is no significative energy
transfer from the homogeneous mode to the fluid, either through parametric
amplification or otherwise. The plots of total particle density and
effective masses presented in ref. \cite{paper2b} suggest that this
condition is met early, then follows an intermediate stage dominated by
rescattering, and finally the thermalization stage.

Comparison with the results in papers 1-5 is meaningful only in the
intermediate phase. Decaying turbulence is necessarily a transient
phenomenon. As time evolves, we expect the field will eventually thermalize,
and the spectrum will get closer to a Rayleigh-Jeans tail, $n_k\sim k^{-1}$
when masses are negligible. The several plots presented in ref. \cite
{paper2b}, where indexes go from $1.7$ to slightly over $1$, capture the
transition from turbulence to equilibrium. Since the same plots show that
earlier spectra are steeper (see also Fig. 1 in ref. \cite{paper2}) this is
in satisfactory agreement with the prediction from self-similar flows.

\epsfxsize 4in
\begin{figure}[htpb]
\centering\leavevmode
\epsfbox{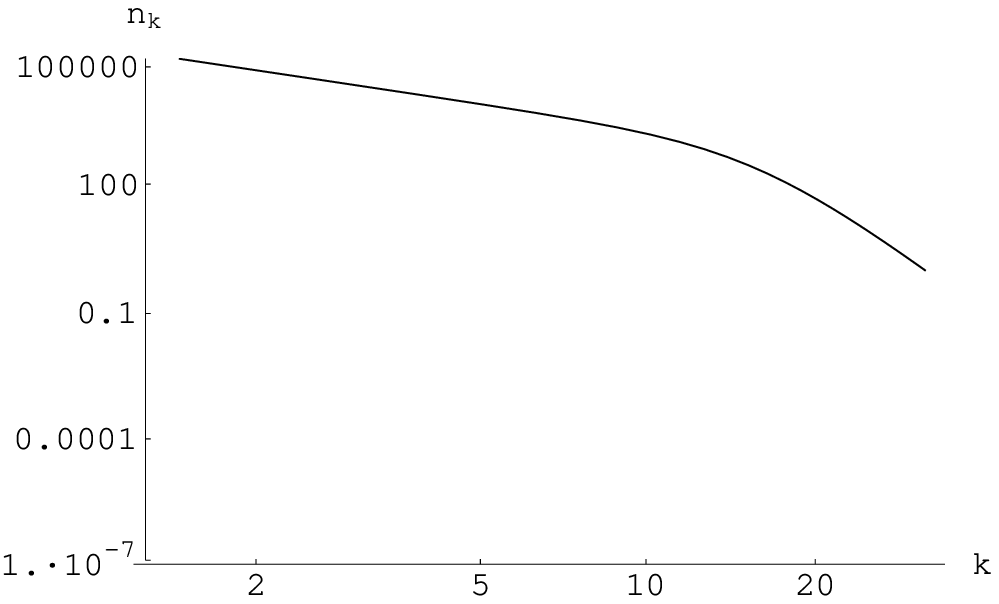}
\end{figure}

Fig. 1: Log-Log plot of the particle spectrum, as given by
eq. (\ref{fluid2quantum}). The energy spectrum is given by (\ref{speccur1}),
where the function $F$ is given by eq. (\ref{fapprox}). We have chosen $A=3$, the Taylor microscale $\lambda \sim 0.18\times 10^{-13}GeV^{-1}$, and have
scaled the spectrum to make it easiest to compare against the results
presented in \cite{paper2} and \cite{paper2b}.

\section{Final remarks}

In this paper, we have shown that the self similar flows studied by
Heisenberg, Chandrasekhar and Tomita may be used to provide an
interpretation of the ''turbulent ''spectra found in papers 1-5. The
hydrodynamic model predicts scale invariant spectra $n_{k}\sim k^{-\alpha }$
both in the infrared and ultraviolet limits, with $\alpha \sim 2$ in the
former, and $10$ in the latter regime. Agreement with the early time results
presented in papers 1-5 is satisfactory.

The connection of hydrodynamics to the behavior of fluctuations during
reheating has interest of its own, as it provides an alternative to brute
force quantum field theoretic calculations, and also yields physical insight
on the macroscopic behavior of quantum fields in the Early universe. The
equivalent fluid method may be used to advantage also in other regimes, such
as the inflationary period itself \cite{Mariana}. Moreover, it opens up a
wealth of new phenomena, such as intermittence \cite{UFrisch} and shocks 
\cite{Burgers}, which are not apparent in the customary treatments. We will
continue our research in this field, which promises a most rewarding
dialogue between cosmology, astrophysics, and nonlinear physics at large.

\section{Acknowledgments}

We thank Mario Castagnino for rekindling our interest in the subject.

EC has been partially supported by Universidad de Buenos Aires, CONICET,
ANPCyT under project PICT-99 03-05229 and Fundaci\'{o}n Antorchas. MG has
been partially supported by National Science Foundation grant PHY97-22022.

\section{Appendix: estimate for the matter viscosity}

In order to estimate the viscosity for the matter fields we assume that the
correspondence eq. (\ref{fluid2quantum}) between the fluid and particle
spectra allows us to associate a solution of the hydrodynamical evolution
equations for the former to a self - similar solution of the kinetic
equation for the latter (cfr. \cite{Zakharov}). We may then estimate the
transport coefficients by adopting the same methods usually applied in
equilibrium \cite{classics}. These are discussed in detail, in the quantum
field theory context, in references \cite{Jeon}.

Let $n_{k}^{0}\sim Nf\left( X\right) $, $X=U^{\mu }k_{\mu }/k_{0}$ be a
solution to the covariant Boltzmann equation

\begin{equation}
k^{\mu }\nabla _{\mu }n_{k}=I_{col}\left[ n_{k}\right]
\end{equation}

Since we are interested in the high energy regime where shear viscosity is
much larger than bulk viscosity, we may use the Boltzmann collision integral

\begin{eqnarray}
I_{col}\left[ n_{k1}\right] &\sim &g^{2}\int Dk_{2}Dk_{3}Dk_{4}\;\delta
\left( k_{1}+k_{2}-k_{3}-k_{4}\right)  \nonumber \\
&&\left\{ \left( 1+n_{k1}\right) \left( 1+n_{k2}\right) n_{k3}n_{k4}-\left(
1+n_{k3}\right) \left( 1+n_{k4}\right) n_{k1}n_{k2}\right\}
\end{eqnarray}

where $Dk\sim d^{4}k\;\delta \left( k^{2}\right) $. To estimate the shear
viscosity we assume the four-velocity $U^{\mu }$ is slightly inhomogeneous,
and seek a solution

\begin{equation}
n_{k}\sim n_{k}^{0}+n_{k}^{1}
\end{equation}

where the new term satisfies the equation (which we render only
schematically)

\begin{equation}
\frac{N}{k_{0}}\frac{df}{dX}k^{i}k^{j}U_{i,j}\sim \frac{\delta I_{col}}{%
\delta n_{k}}\left[ n_{k}^{0}\right] n_{k}^{1}
\end{equation}

By simple power counting, we estimate

\begin{equation}
\frac{\delta I_{col}}{\delta n_{k}}\left[ n_{k}^{0}\right] \sim
g^{2}N^{2}k_{0}^{2}
\end{equation}

And so

\begin{equation}
n_{k}^{1}\sim \frac{1}{g^{2}Nk_{0}^{3}}\frac{df}{dX}k^{i}k^{j}U_{i,j}
\end{equation}

The new term induces a correction to the energy - momentum tensor

\begin{equation}
T_{ij}^{1}=\int Dk\;k_{i}k_{j}n_{k}^{1}
\end{equation}

Equating this to $\eta U_{i,j}$ and repeating our power counting analysis,
we obtain

\begin{equation}
\eta \sim \frac{k_{0}^{3}}{Ng^{2}}
\end{equation}

as in eq. (\ref{noneqvisc}). In equilibrium, $k_{0}\rightarrow T$, $%
N\rightarrow 1$, and we obtain the same result as in refs. \cite{Jeon}.

\end{document}